\documentclass{sig-alternate-05-2015}
\usepackage{graphicx}
\usepackage{amsmath}
\usepackage{todonotes}
\usepackage{paralist}
\begin{document}

%\setcopyright{acmcopyright}
%\doi{10.475/123_4}
%\isbn{123-4567-24-567/08/06}
%\conferenceinfo{}{}
%\acmPrice{}

%\title{Measuring Maturity: Assessing Organizational Security Practices From 
%the Outside }
%\title{Sharing Makes You Sick: Assessing Risk Vectors for Botnet
%Infections}
% Sf not thrilled with this either. 
%\title{Assessing the security risk of an organization through external
%measurements}
%Something Cute: A Data-driven Approach to Understanding Organizational
%Cybersecurity Risk\\or\\Understanding Organizational Cybersecurity Risk:A Data-driven
%approach}
%\title{Outside-In: Externally Assessing the Security Risk of an Organization}
\title{Risky Business: Assessing Security with External Measurements}

\numberofauthors{3}
\author{
  \alignauthor
  Benjamin Edwards\thanks{Work down while at the University of New Mexico} \\
  \affaddr{Cyentia}\\
  \email{ben@cyentia.com}
  \alignauthor
  Jay Jacobs\thanks{Work done while at BitSight} \\
  \affaddr{BitSight}\\
  \email{jay@beechplane.com}
  \alignauthor
  Stephanie Forrest\thanks{Work done while at the University of New Mexico} \\
  \affaddr{Arizona State University}\\
  \affaddr{Tempe, AZ}\\
  \affaddr{Santa Fe Institute}\\
  \affaddr{Santa Fe, NM}\\
  \email{steph@asu.edu}
}

%\author{\IEEEauthorblockN{Benjamin Edwards}
%\IEEEauthorblockA{Dept.\ of Computer Science\\
%University of New Mexico\\
%Albuquerque, NM}
%\and
%\IEEEauthorblockN{Jay Jacobs}
%\IEEEauthorblockA{BitSight Technologies\\
%Cambridge, MA\\}
%\and
%\IEEEauthorblockN{Stephanie Forrest}
%\IEEEauthorblockA{Dept.\ of Computer Science\\
%University of New Mexico\\
%Albuquerque, NM\\
%Santa Fe Institute\\
%Santa Fe, NM}}

\maketitle

\begin{abstract}

Security practices in large organizations are notoriously difficult to assess.
The challenge only increases when organizations turn to third parties to
provide technology and business services, which typically require tight network
integration and sharing of confidential data, potentially increasing the
organization's attack surface. The security \emph{maturity} of an
organization describes how well it mitigates known risks and responds to new
threats. Today, maturity is typically assessed with audits and
questionnaires, which are difficult to quantify, lack objectivity, and may not
reflect current threats.

This paper demonstrates how external measurement of an organization can be used
to assess the relative quality of security among organizations.  Using a large
dataset from BitSight\footnote{\texttt{https://www.bitsight.com}}, a cybersecurity ratings company, containing 3.2 billion
measurements spanning nearly 37,000 organizations collected during calendar
year 2015, we show how per-organizational `risk vectors' can be constructed
that may be related to an organization's overall security posture, or maturity.
Using statistical analysis, we then study the correlation between the risk
vectors and botnet infections.  For example, we find that misconfigured TLS
services, publicly available unsecured protocols, and the use of peer-to-peer
file sharing correlate with organizations that have increased rates of botnet
infections. We argue that the methodology used to identify these correlations
can easily be applied to other data to provide a growing picture of
organizational security using external measurement. 
%The analysis identifies statistical correlations rather than analyzing direct
%causation, and only gives a partial picture of maturity of organizations.
%However, our methodology widely applied to new information to give a more the
%relative security maturity of an organization. 

\end{abstract}

\section{Introduction}
\label{sec:intro}

%In 2015, there were 6,453 software vulnerabilities published in the National
%Institute of Standards and Technology's(NIST) National Vulnerability
%Database~\cite{NVD}. Given the constant stream of new threats to organizational
%networks, it is crucial that information security practitioners have systematic
%and robust methods for keeping enterprise networks functioning and free from
%malware. Measuring how well an organization handles new and existing security
%threats is a challenge. This question is increasingly relevant as organizations
%rely more on their technical infrastructure for everyday tasks.
%
%This question has traditionally been answered through audits and penetration
%testing~\cite{arkin2005software}. However, many organizations do not invest in
%security standards compliance, and most are unaware of which of the many
%standards in existence with which they should strive to comply~\cite{ftse2013}
%Moreover, while these types of investigations are able to identify potentially
%dangerous practices and vulnerabilities, they do not provide a direct measure
%of the maturity of an organization. In particular, they do not provide a
%measure of the degree to which the presence of vulnerable systems and software
%are correlated with negative outcomes like botnet infections.

In an increasingly connected world, organizations frequently partner with third
parties. A 2014 report by the Institute for Internal Auditors found that 91\%
of organizations partnered with technology vendors, 76\% with business service
providers, and 40\% formed strategic partnerships with third
parties~\cite{warren2014closing}. These partnerships often entail sharing
confidential data and integrating network infrastructure, potentially
increasing an organization's attack surface and risk. One high profile example
was the 2014 Target breach, which exposed credit card information about 40
million customers~\cite{harris2014target} when attackers stole a third
party vendor's credentials and used them to infiltrate the
network~\cite{perlroth2014heat}. 

Traditional approaches to managing the risk in such partnerships use risk
assessment
questionnaires~\cite{assessments2010standardized,brock1999information} and
audits~\cite{council20142015}.  The results are then interpreted with cyber
threat matrices~\cite{mateski2012cyber}, which provide a rough, qualitative
snapshot of risk. Such strategies are time consuming, expensive, and it is
unclear how effective they are.  Although some efforts have been made at
standardization~\cite{assessments2010standardized}, there is currently no
standard quantitative and objective approach to assessing risk in these
environments. 
% SF felt that this next sentence detracted from our overall point.
%Risk assessment is further complicated because many organizations do not
%invest in security standards compliance, and most are unaware of which of the
%many exisitng standards are most relevant to their
%organization~\cite{ftse2013}.  
Finally, although assessments, audits, and compliance standards can point to
vulnerabilities (potential avenues of attack), they do not link these
vulnerabilities to actual outcomes.  Thus, there is a need for a different
approach to the problem of understanding and mitigating security risks in
organizations.  Ideally, such an approach will be objective (not subject to
self-interpretation), noninvasive, quantitative, and it will reflect actual
risk, rather than hypothetical threats.

This paper addresses the need by presenting a rigorous, data-driven methodology
for assessing organizational risk vectors. The methodology can inform an
organization about the risks posed by third-party partnerships, and it can help
it better understand its own risk profile, ultimately providing guidance on how
to improve the security of its internal networks. The methodology consists of
three components: 1)a mapping of network (IP) space to individual organizations
2) measurement of possible avenues of attack which we dub `risk vectors' and 3)
measurement of externally observable security incidents.

We focus on risk vectors that can be measured externally and objectively and
show how  they correlate with actual security incidents. We investigate three
broad classes of risk vectors: peer-to-peer file sharing, incorrect
configuration of Transport Layer Security (TLS) services, and the presence of
publicly available insecure communication protocols. For simplicity, we focus
on assessing risk using one type of security incident, the presence and
prevalence of botnet infections within an organization. While there are myriad
possible security incidents to measure, botnets are a critical part of the
cybercrime infrastructure, and cost users upwards of \$10 billion in cleanup
costs alone~\cite{anderson2013measuring}.  Because our method is statistical in
nature, it doesn't identify root causes of any particular security incident.
Rather, it provides an overall assessment of the security \emph{maturity} of an
organization, in the same spirit as assessments and audits.

To investigate the link between risk vectors and botnet infections, we begin
with a large dataset consisting of over 3.2 billion network events measured
across almost 37,000 organizations throughout 2015 (Section~\ref{sec:data} C).
IP addresses are associated with specific organizations through a rigorous and
unique mapping process (Section~\ref{sec:data} A). The mapping allows us to use
large-scale scans of the Internet and associate specific kinds of events with a
particular organization.  Combining these data with information about an
organization's size and other properties (Section~\ref{sec:data} B) allows us
to normalize measures of risk and botnet infection, study different types of
organizations, and make comparisons across similar organizations.
%We find that 90\% of organizations have fewer than 1 botnet infection per 12
%employees, however, infection rates can span many orders of magnitude.

With these data in hand, we show how they can be leveraged to develop
statistical models to establish quantitative relationships between risk vectors
and botnet infections. We find that each of the three categories of risk vector
correlates with the presence and prevalence of botnets within an organization. 
% Details moved to conclusion
We also study how the effect\footnote{We use the term \emph{effect} in the
statistical sense to indicate the quantitative magnitude of a
phenomenon~\cite{kelley2012effect}. Specifically we do not mean it to imply a
\emph{causal} relationship.} of each risk vector varies with the type of
organization, finding significant differences between different organization
types. 

We do not claim that the link between risk vectors and botnet infections is
causal. However, we do suggest that they have a common cause---security
immaturity.  Failure to configure TLS services correctly,
making known insecure protocols publicly available, and allowing
employees to download files that are likely infected with malware, are all
indicative of poor network security practices. Similarly,
the failure of an organization to detect and remove botnet infections suggests
that it probably lacks a systematic and thorough approach to security.  In this
way, the paper shows how security maturity
%The paper shows how both of these 
can be effectively assessed through external measurements of risk vectors and
infections.
% Even though our results are based on a very small sample of all possible risk
% vectors and admitting that our measurements  may obscure important technical
% details, our results suggest that  noninvasive and objective assessments of
% organizational maturity are feasible
%can act as a compass for traditional methods such as risk assessment
%questionnaires, audits, and penetration testing. Moreover, 

Our results only focuses on a handful of risk vectors which
are externally observable, and one type of security incident. However, our
methodology is general and could easily be expanded to more risk vectors and
different security incidents, e.g.\ data breaches and service interruptions. We
believe that moving towards a more quantitative data-driven view of risk
assessment will be crucial in the future. Not only will provide a
clearer picture of cyber risk, but will help to assess whether specific
security lapses are associated with incidents. This in turn will help security
practitioners assess and triage emerging threats for their specific
organization.
%For example, our approach could be used to identify which risk vectors are
%associated with security incidents that have a larger financial impact such as
%data breaches or service interruptions.
In summary the paper makes several contributions:

\begin{compactenum}

  \item A methodology for collecting security data about organizations using
    external measurements.

  \item A demonstration of how to analyze the collected data to gain insight
    about the relative security of organizations

  \item Insights from the data including: a strong relationship between
    peer-to-peer sharing activity and botnet infections, and differentiation of
    risk vectors across different industries.

  \item A thorough discussion of how the methodology could be expanded and used
    to cover other types of incidents and risk vectors.

\end{compactenum}

The rest of the paper is organized as follows: Section~\ref{sec:data}
elaborates on our process for data collection, organization, and aggregation.
Next section~\ref{sec:analysis} provides an initial analysis of the data and
examines some of it's basic properties. Section~\ref{sec:models} builds several
models of risk vectors and botnet infections. Section~\ref{sec:related} cover
related work. We conclude with a discussion of the implications of our results
and opportunities for future work in section~\ref{sec:discussion}, and some
final remarks in section~\ref{sec:conclusion}.

\section{Data Collection and Processing}
\label{sec:data}

This section describes the data collection and processing that were used in
this study. The dataset was collected throughout 2015 by BitSight for use in
their commercial service. We discuss how the data was used to develop the
mapping of IP space to organizations, the measurement of risk vectors, and the
measurement of security outcomes.

%\begin{figure}[ht!]
%  \includegraphics[width=\columnwidth]{./figs/placeholder.png}
%  \caption{A flow chart demonstrating the process for mapping raw data streams
%  into usable data}
%  \label{fig:flowchart}
%\end{figure}
  
\subsection{Identifying an Organizations Network Footprint}

The first step to measuring an organizations security practices is to identify
its overall network presence. Over the past four years, BitSight has developed
a process for identifying the IP addresses associated with individual
organizations, which they use for in their commercial service. In this section
we give an overview of our process. A variety of methods could be used to
construct a mapping of IP space to organizations, but it is crucial to make
this mapping as accurate as possible. Given the heavy tailed nature of security
incidents~\cite{edwards2016hype,edwards2015analyzing}, misc attribution of a security incident could
alter the assessment of an organization.

For this reason the process presented here utilizes a manual verification step.
Specifically, individual researchers construct mappings for each organization,
which is then independently checked by another researcher. This process
presented here is an improvement over the one used in~\cite{liu2015cloudy} as
it uses additional information sources and completed and verified manually. 

Organizations are selected because they are economically prominent (such as the
Fortune 500 companies) or are referred for mapping through industry partners. A
researcher starts the mapping process by first identifying an organization's
web presence and main company domain. Next, a company's presence on social
media as well as other public sources are examined and information about the
company is gathered. This includes the industry in which the organization
operates and its number of employees.  The number of employees gives a rough
estimate of the organization's size, which we use to normalize some of our
other measurements so meaningful comparisons between different organizations
can be made. This is an imperfect measure of size however, and we discuss some
consequences in section~\ref{sec:analysis}, and suggest how it might be
improved in~\ref{sec:discussion}. Next, media sources and financial fillings
are used to identify any fully owned subsidiaries of an organization.  After
this organizational information is gathered, a manual search of BGP routing
information, regional internet registries, and other proprietary services are
consulted to identify IP addresses which are allocated to that organization and
its subsidiaries.

We believe this represents a best-effort approach to accurately identifying
organizational ownership of IP addresses and thus observed events and services.
At the time these data were collected, BitSight has mapped 1.8 billion IPv4
addresses (42\% of the IPv4 Internet), enabling us to observe 36,982 distinct
organizational entities in 2015. This includes all of the Fortune 500
companies. However, any methodology which endeavors to associate IP addresses
to organization will face many challenges, such as the deployment of DHCP and
NAT and the increasing use of cloud services. We believe our approach improves
on previous similar mapping techniques~\cite{liu2015cloudy} by consulting
multiple additional sources, manual identification of IP ranges, and includes a
manual verification step.

%\subsection{Mapping IP addresses to Organizations}
%\label{subsec:mapping}
%
%Over the past four years, BitSight has developed a mapping from IP addresses to
%specific organizations, which they use in their commercial service.  The
%mapping incorporates data sources such as BGP routing information and
%regional registries, which is then verified manually. 
%At the time these data were collected, BitSight had 1.8 billion IPv4 addresses
%(42\% of the IPv4  Internet) mapped, enabling us to observe 36,982 distinct
%organizational entities with events in 2015.
%
%\subsection{Organization Properties}
%
%In addition to the IP addresses associated with each organization we also
%collect some descriptive information. During the manual mapping process a
%variety of public sources were examined to discover information about the
%company. This includes the organization's industry categorization and its
%number of employees. The number of employees gives a rough estimate of the
%organization's size, which we use to normalize our results allowing meaningful
%comparisons between different organizations. This is an imperfect measure
%however, and we discuss some consequences in section~\ref{sec:analysis}, and
%suggest how it might be improved in~\ref{sec:discussion}.

\subsection{Risk Vectors}

Here we describe a handful of externally observable events and system states,
which we refer to as ``risk vectors'' (also known as risk factors). Most are
unlikely to directly cause malware infections in and of themselves (though
peer-to-peer activity might be an exception); rather, they are indicators of
conditions in an organization that may lead to malware infection or other
security problems.  That is, risk vectors relate to an organization's security
maturity.  We consider three classes of risk vectors: peer-to-peer file
sharing, transport layer security, and network services.
 % Each vector (sometimes referred to as a risk factor) represents an
% opportunity for the organization to better secure their infrastructure, and
% therefore reduce the attack surface of their network.  We further describe
% how we scale each individual risk vector so that comparisons can be made
% across organizations. 

\subsubsection{Peer-to-Peer File Sharing}

Peer-to-peer file sharing protocols are well-known security risks.
% Beyond the potential for liability from downloading copyright
%material, 
Research has shown that as many as 35\% of torrent files are infected with
malware~\cite{cuevas2014torrentguard}, and the infrastructure itself can be
used to propagate worms~\cite{hatahet2010new}.  Since peer-to-peer file sharing
has only limited use in most enterprises, many security-conscious organizations
block BitTorrent by default within their networks or prevent their users from
installing and running torrent clients. While other peer-to-peer file sharing
protocols exist we focus on BitTorrent because it is the most popular
protocol~\cite{sandvine2015global}.
%
% . This is not a trivial endeavor, because the protocol is flexible and
% without Deep Packet Inspection (DPI) it is difficult to identify torrent
% traffic~\cite{mueller2012deep}.  Therefore, manyHowever, in an enterprise
% environment simply preventing the installation of torrent clients should be
% sufficient to keep employees from using torrent protocols.

We identify peer-to-peer file sharing by collecting torrent tracker lists from
%the two largest open, public torrent trackers, The Pirate
%Bay\footnote{https://thepiratebay.org} and Kick Ass
%Torrents\footnote{https://kat.cr}.
two of the largest open and public trackers. 
%We connected to each torrent and recording the IP address associated with each
%file share.  
For each torrent in the tracker list we collect a list of IP addresses are
`seeding' the file, meaning the IP addresses that have downloaded the complete
file and made it available for download by others. The IPs are then mapped to
organizations as described above.  This provides a count of the number of files
actively shared by each organization, which we normalize by the number of
employees to produce a per employee concentration. By counting the total number
of files rather than IP addresses sharing, we believe we obtain a more accurate
representation of the prevalence of file sharing than would be obtained by raw
IP counts.

\subsubsection{Transport Layer Security}
\label{subsec:TLS}

Transport Layer Security (TLS) is the backbone of encrypted internet
communication and is often the target for new attacks and vulnerabilities.

We track two different types of possible errors in the TLS protocol, software
configuration weaknesses and certificate weaknesses.  Software weaknesses are
caused either by out-of-date software or an administrative error in the
configuration of the service. Out-of-date software leaves the network
susceptible known vulnerabilities such as
Heartbleed~\cite{durumeric2014matter} and FREAK~\cite{beurdouche2015messy},
while the use of weak versions of the Diffie-Helman key exchange could lead to
eavesdropping~\cite{adrian2015imperfect}. 
%Misconfigurations can also lead to the use of vulnerable cryptographic
%algorithms. such as allowing the use of outdated versions of the protocol such
%as SSLv2 and SSLv3, 
It is unlikely that TLS weaknesses would lead directly to malware infections at
any measurable scale.  More commonly, the result is  
%The most common result of TLS weaknesses is a loss of
that some data are no longer communicated confidentially.  However, TLS issues
provide an excellent indicator of the state of an organization's security
maturity.
% in service, not the integrity of the system hosting the service
%(which would enable malware infections). For example, The Heartbleed
%vulnerability, however, can cause unrestricted access to the vulnerable
%server's active memory, potentially leading to the compromise of
%credentials~\cite{durumeric2014matter}.  Moreover, some of the attacks are
%purely hypothestical, for example in~\cite{adrian2015imperfect} the authors
%state that they believe that it is plausible that systems using 1024-bit key
%exchange would be vulnerable to an attacker with nation-state resources, but
%prevent no evidence that such an attack exists. We address this link more in
%section~\ref{sec:discussion}.

We derive data on the number, type, and configuration of TLS services from
Internet wide scans of IPv4. These scans probe all of IPv4 space and attempt to
identify any running services across a number of ports utilized for common
services~\cite{durumeric2013zmap}. 250 different ports commonly used for a
variety of services were scanned. We limit our investigations to ports offering
TLS and 21 ports commonly associated with popular services (see
section~\ref{subsec:services}). 
%These scans were obtained from a popular Internet scanning search engine.
Scans were completed roughly once per month, and if a TLS service was present,
a connection was established with the server, and the certificate presented as
part of the process was saved for analysis.
%It would be ideal to record all of the possible TLS services running at each
%organization exactly once per month month. 
While faster scanning processes exist~\cite{durumeric2013zmap}, our approach
utilizes more in depth scanning, for example it tests all possible encryption
suites and protocol versions when establishing a TLS connection. This less
frequent scanning also reduces the chance that the scans will be perceived as
malicious and blocked.
%However, our scanning process is not that precise.  is probable that all
%services for each organization were not scanned on a monthly basis. 
Since we find little variance in the total number of TLS services for each
organization each month, we believe that the scans are accurate enough for our
purpose.

Certificate errors refer to problems with the certificates used for
authentication and the Public-Key Infrastructure. The collected certificates
were examined to determine if they used keys created using weak cryptographic
protocols, were signed using cryptographically weak hash functions, or had a
suspicious chain of trust. 
% For example it has been demonstrated that certificates that utilize 512-bit
% RSA have been exploited in several attacks~\cite{FoxIT2011}.  Similarly, it
% has been shown that collision attacks are effective against
% SHA1~\cite{wang2005finding}, MD5~\cite{dobbertin1996status}, and
% MD2~\cite{rogier1997md2}, indicating that none should be used to sign
% certificates as fake certificates with the same signature could be generated
% by attackers. 
Specifically, we collect information on chain of trust issues including expired
certificates, certificates that were issued for a future date, self signed
certificates, and certificates whose chain of trust is broken. Any of these
errors alone could be benign, but all are potentially exploitable by a
determined attacker and relatively easy to fix, and therefore a good candidate
indicator of the maturity of the organization. 

%Evidence for this is that there is larger variance than in month
%to month SSL service counts than we would expect from organizations adding and
%dropping such services.\footnote{Not Shown.}

Specifically, we looked for the following errors in software configuration:
\begin{compactitem}
  \item TLS version less than or equal to SSLv3;
  \item The presence the Heartbleed or FREAK bugs;
  \item The presence of weak Diffie-Helman Key exchange, either keys with
    less than 2048 bits or with commonly used prime numbers;
\end{compactitem}

\noindent and certificate errors, including:
\begin{compactitem}
  \item Self signed certificates, 
    expired certificates, certificates that were issued in the
    future, certificates with nonstandard roots,
    and certificates with a broken chain of trust;
  \item Certificates with weak keys, specifically using RSA or DSA with 1024 
    bits or less, or ECC with less than 224 bits;
  \item Certificates with weak signatures, including those signed with SHA1, MD5,
    and MD2.
\end{compactitem}

Larger organizations are likely to have a higher number of TLS services.  To be
able to make comparisons across organizations, we calculate the fraction of TLS
services that have configuration and certificate errors. Each of these
fractions is used as a separate risk vector in our analysis.

\subsubsection{Services}
\label{subsec:services}

Finally, we measure how many and what kind of network services the organization
makes publicly available. While scanning for TLS services we also scanned for
other types of services on other ports. Although any external communication
service increases the attack surface of an organization, some services are
safer than others. For example, remote terminal access through SSH is encrypted
and supports key-authentication, so it is preferred over terminal access
through Telnet, which transmits all data through plain text, leaving an
organization susceptible to eavesdropping and interception of passwords and
data~\cite{wagner2001address}.

We scanned 250 different services and categorized 21 frequently used services
into \emph{risky}, \emph{neutral}, and \emph{reasonable}. These services
account for just under 95\% of all services seen in the scans and were selected
because they were popular or posed significant security risks.  Reasonable
services are those that can be exposed to the Internet with relative little
fear of easy exploit such as SSH, while risky services should not be open to
public access(telnet). Neutral services are those which may be exploited given
misconfiguration or unpatched software, but are not inherently dangerous or are
required to be exposed to the outside world to be useful, for example HTTP.\
These services are summarized in table~\ref{tbl:services}.

\begin{table}
\caption{Classification of publicly available services.}
  \label{tbl:services}
  \centering
  \begin{tabular}{|l|l|l|}
    \hline
    \textbf{Service} & \textbf{Class} & \textbf{Classification Reason}\\
    \hline
    FTP & Risky & Clear text communication \\
    \hline
    TELNET & Risky & Clear text communication\\
    \hline
    SMTP & Risky & Clear text communication\\
    \hline
    POP3 & Risky & Clear text communication\\
    \hline
    SUNRPC & Risky & Multiple vulnerabilities~\cite{sunrpc}\\
    \hline
    NETBIOS & Risky & Network foot-printing~\cite{netbios} \\
    \hline
    IMAP & Risky & Clear text communication\\
    \hline
    SNMP & Risky & Multiple vulnerabilities~\cite{snmp}\\
    \hline
    SMB & Risky & Multiple vulnerabilities~\cite{smb}\\
    \hline
    MYSQL & Risky & Direct database access\\
    \hline
    MSSQL & Risky & Direct database access\\
    \hline
    RDP & Risky & Multiple vulnerabilities\\
    \hline
    POSTGRES & Risky & Direct database access\\
    \hline
    DNS & Neutral & Vulnerable but necessary\\
    \hline
    HTTP & Neutral & Vulnerable but necessary\\
    \hline
    NTP & Neutral & Vulnerable but necessary\\
    \hline
    SSH & Reasonable & Encrypted\\
    \hline
    HTTPS & Reasonable & Encrypted\\
    \hline
    SMTPS & Reasonable & Encrypted\\
    \hline
    IMAPS & Reasonable & Encrypted\\
    \hline
    POP3S & Reasonable & Encrypted\\
    \hline
  \end{tabular}
\end{table}

For each organization we normalize by calculating the fraction of known
services that were classified as reasonable, neutral and risky. 

\subsection{Botnet Infections}
\label{subsec:bots}

The above measures identify the network properties of an organization, but do
not directly measure security incidents. In this paper we examine a common
security incident: botnet infections. In this section, we describe how data
from a diverse number of botnets is collected as a measure of security
incidents within a company. 

Botnet infection data is collected through Anubis Networks. Anubis Networks
uses several techniques to infiltrate botnet command and control structure so
measurements of individual infections can be taken. First, samples of malware
are obtained and are reverse engineered to identify how a particular bot
communicates with its command and control infrastructure. Many botnets
communicate through randomly generated domains, created using Domain Generating
Algorithms(DGA) which are controlled by the botnets
master~\cite{porras2009analysis}. By reverse engineering the DGA used to create
these random domains, Anubis is able to register domains consistent with the
algorithm. Communications from the infected client to these domains allows for
the monitoring of IPs associated with infections~\cite{dagon2006modeling}.

Some botnets have attempted to circumvent this type of infiltration by using a
peer-to-peer architecture to communicate with the botnet
master~\cite{grizzard2007peer}. In these cases, a machine controlled by Anubis
executes the malware and communication with other members of the botnet is
monitored. These bot peers can then be queried to further investigate the
peer-to-peer structure.

Using this methodology, the activity of 120 different botnets was monitored in
2015. In this time period over 650 million unique infection days were
identified, where an infection day is a unique IP address infected on a
specific day. For each organization we measure this count of infection days on
a monthly basis. For example, if an organization has a single IP address which
is infected for a single day in a month, its infection day count is 1. Whereas
if that single IP address is infected for a week its infection day count is 7.
This measure allows us to measure the severity of infection by both the total
number of infected IP addresses and the duration of infections. We then
normalize these counts by employee count to obtain a concentration of infection
days per employee per month. Because of the presence of NATs and DHCP, these
measurements represent a lower bound on infection concentrations. We discuss
how we might address this in the future in section~\ref{sec:discussion}. 

%These sinkholes experienced billions of connections over the course of 2015.
%After eliminating duplicates we identified nearly 650 million unique systems
%compromised by botnet infections during the period.
%These communications were carefully aggregated to individual IP address.
%These data were then used to identify the IP addresses that are participating
%in each botnet. The IP addresses were then mapped to organizations (see above)
%to identify the count of unique addresses experiencing an infection by an
%organization each month. Finally, we normalize the count, as before, to obtain
%the botnet infection rate per employee.

\subsection{Data Aggregation}

While it is conceivable that some of these data could be collected
continuously, for example the botnet data could be collected in real time, most
collection requires longer time scales. Scanning available services requires
roughly a month to scan all the possible IPs associated with the organizations
we are monitoring.  Therefore, we aggregate the data by month,
% obtain monthly aggregate counts for each entity
%of the network events below. Aggregating and normalizing each of the
%quantities
producing monthly counts and rates for each organization.  We found little
variation in any of the measures over time, so we averaged the values of the
measured quantities over the entire year.
%and in order to obtain a more general idea of effect of risk vectors on botnet
%infections, we took the average values of the quantities over the course of
%the year. 
Future research could investigate the impact of time, for example, how risk
vectors and infections change after major security incidents such as data
breaches.

\section{Analysis of Risk Vectors and Infections}
\label{sec:analysis}

This section examines the data aggregated in~\ref{sec:data}. First, we
investigate some of the basic properties of the data. Next, we ask whether our
risk vectors (peer-to-peer file sharing, TLS errors, and publicly accessible
services), correlate with botnet infections. We find that risk vectors are
correlated with botnet infections in ways we would expect.

\subsection{Risk Vectors}

\begin{figure}[t!]
  \includegraphics[width=\columnwidth]{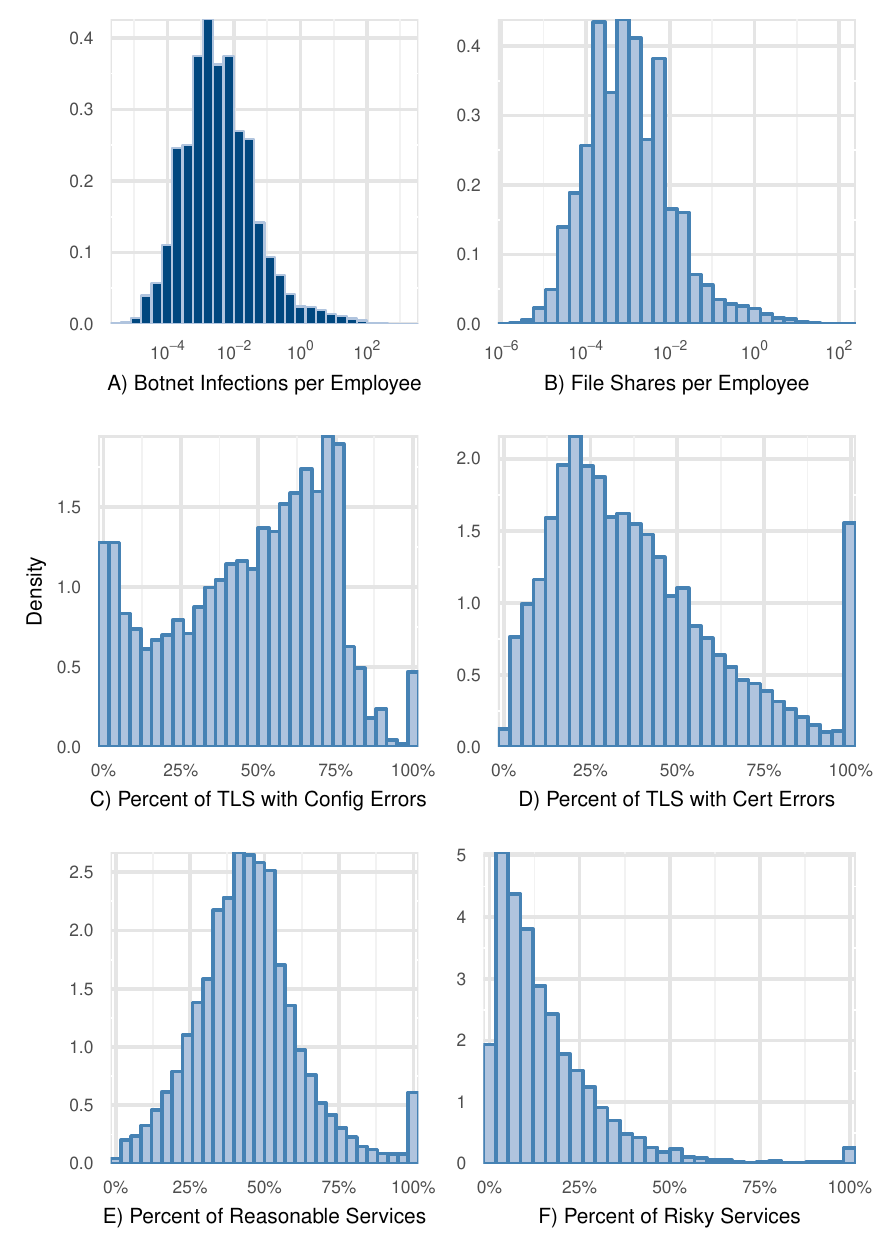}
  \caption{Histograms of risk vectors and botnet infections. Note that plots of
  botnet infections per employee (A) and peer-to-peer file sharing per employee
(B) are plotted on log scales, as they span several orders of magnitude. All
others are on linear scales. (A) appears as a slightly different color as it is
an outcome as opposed to the other variables which are risk vectors.}
  \label{fig:distribution}
\end{figure}

We begin the analysis by visualizing the distribution of each risk vector and
infection variable plotted as histograms in Figure~\ref{fig:distribution}.

Figure~\ref{fig:distribution} demonstrates that organizations that have 
botnet activity, tend to have relatively low levels of botnet infections (90\%
of organizations with observed botnets, have less than one bot per 12
employees). This does not account for the 67\% of organizations that
have no botnet activity at all. However, the data are highly skewed with botnet
infections per employee spanning several orders of magnitude. Peer-to-peer file
sharing has similar properties with the majority of organizations having no
peer-to-peer activity, and among those that do 90\% have less than 0.02 shares
per employee.  In several cases there are more total infections (or shares)
than our estimate of the number of employees. For organizations such
universities or telecommunications companies, this is likely the case, as there
are far more machines with the potential for bots and peer-to-peer sharing than
actual employees. We discuss possible alternative approaches to normalization
in section~\ref{sec:discussion}.

To account for the large number of zeros in the data we analyze both the
\emph{presence} and \emph{prevalence} of botnet infections and peer-to-peer
file sharing in our analysis.
Figure~\ref{fig:distribution} also shows a large percentage of the measured
values of certificate and service errors are either zero or one. That is, for
many organizations either all of their TLS services were misconfigured or none
were.  In between these two values we see a relatively smooth distribution.  A
similar property can be seen in risky and reasonable services, though it is not
as pronounced, with both types of services experiencing relatively high density
at one. We discuss the implications of this type of distribution in
section~\ref{sec:discussion}.

\subsection{Identifying Correlations Between Risk Vectors and Infections}

To establish whether a relationship exists between risk vectors and security
outcomes we use three statistical methods. When both measures are continuous,
we use Spearman's $\rho$, a nonlinear measure of
correlation~\cite{daniel1990applied}. We use Spearman's $\rho$ instead of the
traditionally used Pearson's correlation coefficient, because some of the
relationships may be nonlinear and we would like to capture any dependence
between the values.

When one value is continuous and the other is discrete, for example the
presence the existence of botnets and the fraction of services which employ
encryption, we use Mann-Whitney-Wilcoxon rank-sum text~\cite{mann1947test}.
This test is similar to the Student's $t$-test in that it tests whether one
particular distribution is larger than another, except it does not rely on an
assumption of normality for the underlying distributions~\footnote{We do not
  use the test to identify a difference in medians, which requires stronger
assumptions about underlying shapes of the distributions}. Finally, when both
values are discrete, we use a G-test, which is a measure of statical dependence
in discrete variables. It is recommended that the G-test be used in situations
where the $\chi^2$-test was traditionally used, as the $\chi^2$-test was
conceived as an easily calculable approximation of the likelihood ratio
test~\cite{sokal1969principles}. 

\subsection{Risk Vectors and Infections}

\begin{figure}[t!]
  \includegraphics[width=\columnwidth]{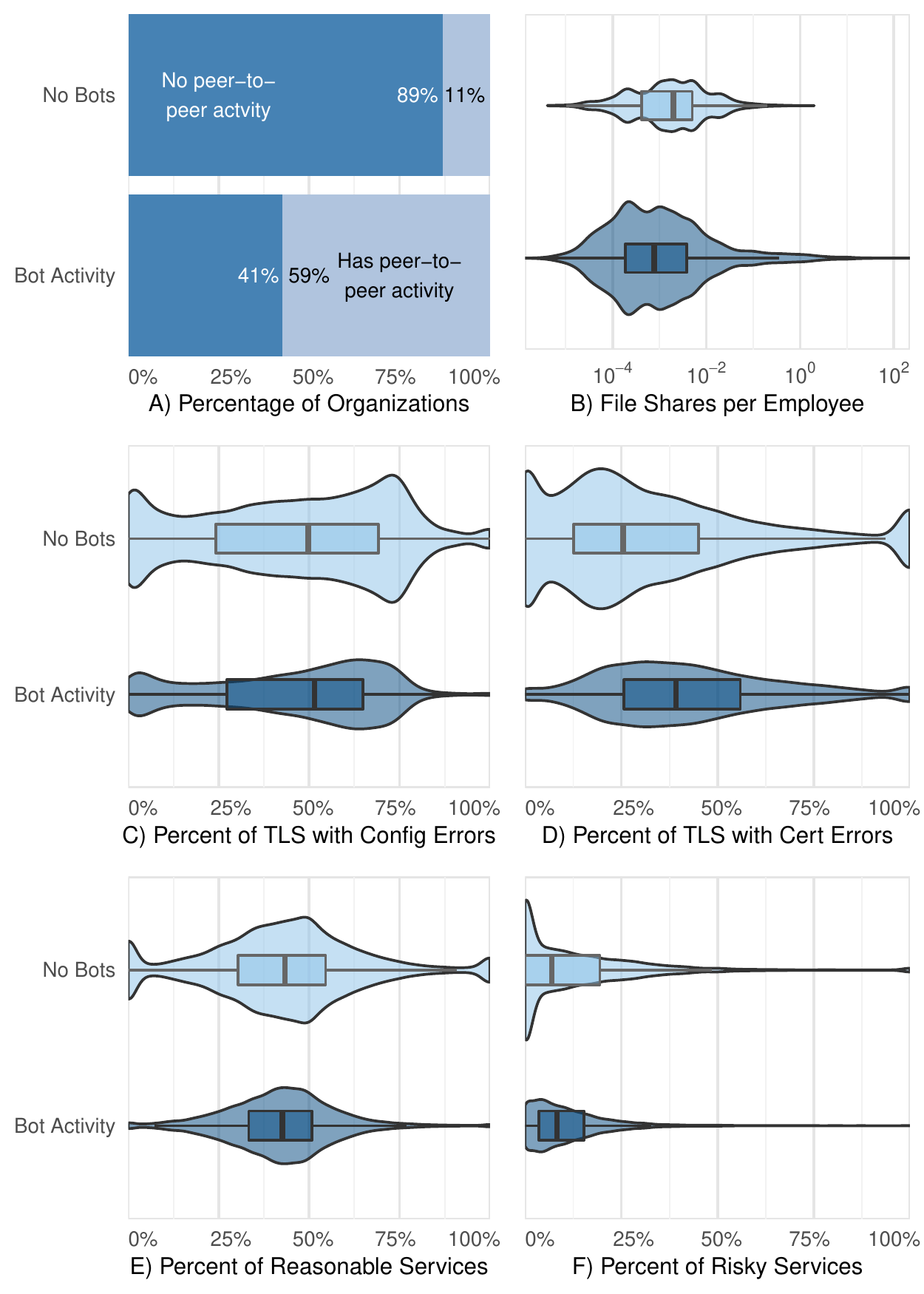}
  \caption{A) shows the percentages of organizations with and without bots and
  the relative number of those organizations that have peer-to-peer activity.
B) C) D) E) and F) are violin plots of the distributions of risk vectors in
organizations measured to have botnet activity and those without botnet
activity. Each subplot contains a violin plot which is a representation of the
underlying distribution of values as represented by a Kernel Density Estimate.
Box and Whisker plots are also included showing the 25\% 50\% and 75\%
quartiles(Box) and 1.5 times the inner quartile range (Whiskers).}
  \label{fig:violin}
\end{figure}

Because a large number of organizations experience no botnet activity, we first
test whether any of the risk vectors are associated with the presence of bots
in an organization. To do this, we consider the distribution of each of the risk
vectors for organizations which we have measured botnet activity and those
which we have not. The results can be see in figure~\ref{fig:violin}.

A very clear relationship can be see in panel A) of figure~\ref{fig:violin}:
organizations with peer-to-peer activity tend to have some botnet activity,
while those without peer-to-peer activity tend to lack botnet
activity($p<10^{-12}$). For the other risk factors the results are not easily
visually discernible; however, we can establish the difference using
statistical tests. For three of the five remaining risk vectors, the
relationship between botnet infections and the risk vector is what we would
expect, organizations with bots generally have higher levels of TLS certificate
errors (D), risky services (E) and lower levels of reasonable services (F).
These differences are statistically significant using the Mann-Whitney-Wilcoxon
rank test ($p<10^{-4}$).

Counter-intuitively the prevalence (shares per employee) is slightly higher for
organizations without bots ($p<10^{-12}$). This is also the case in the
difference in the distribution of TLS configuration errors between
organizations with bots ($p<10^{-2}$), and those without. We discuss possible
explanations for these results in section~\ref{sec:discussion}.

\subsection{Botnet and Risk Vector Correlation}

\begin{figure}[t!]
  \includegraphics[width=\columnwidth]{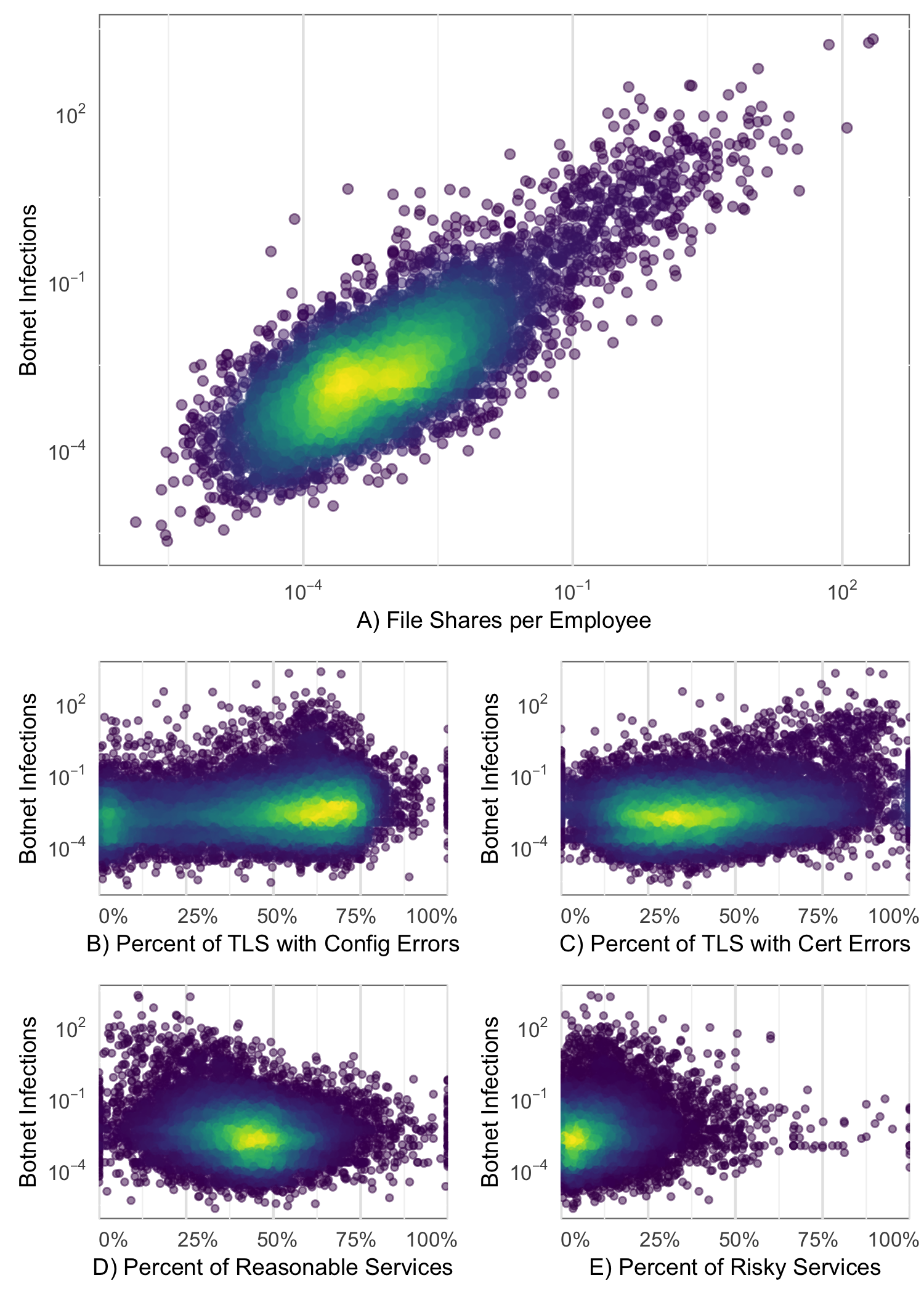}

  \caption{Scatter plot of botnet activity and risk vectors. Colors indicate
  the density of points in an area (lighter colors are higher density, darker
colors are lower density). Densities in range from 0 to 0.041 in A), 0 to 0.36
in B), 0 to 0.35 in C), 0 to 0.59 in D), and 0 to 1.10 in E)}
  
\label{fig:scatter}
\end{figure}

The relationship between the count of IP addresses associated with botnets per
employee and various risk vectors is plotted in Figure~\ref{fig:scatter}.
Values of the Spearman's Correlation Coefficients can be seen in
table~\ref{tbl:coeff}.

We can see that peer-to-peer sharing has a strong positive relationship to
botnet infections. In particular we note that it appears roughly linear on a
log-log scale\footnote{Bivariate linear regression on log transformed variables
revealed a slope of 0.879}. This may indicate the existence of a power-law
relationship. The other variables have weak, but statistically significant
linear relationships to botnet infection rates. 

\begin{table}
  \caption{Correlation Coefficients between risk vectors and botnet count per
  employee. All correlations are significant at the $p<10^{-12}$ level.}
  \label{tbl:coeff}
  \centering
  \begin{tabular}{|c|c|}
    \hline
    \textbf{Variable} & \textbf{Spearman's} $\rho$ \\
    \hline
    Concentration of Peer-to-Peer & 0.725 \\
    \hline
    TLS Configuration Errors & 0.176 \\
    \hline
    TLS Certificate Errors & 0.174\\
    \hline
    Risky Services & 0.138\\
    \hline
    Reasonable Services & -0.151\\
    \hline
  \end{tabular}
\end{table}

The other risk vectors show weak but significant correlations. The negative
correlation between reasonable services and botnet infections may at first seem
counterintuitive, as running any services, even if they are reasonable, would
increase an organizations attack surface, increasing the potential for
infections. However, because we are measuring the fraction of services which
are reasonable, organizations with a high fraction of reasonable services (and
necessarily a lower fraction of risky services), are likely to have a smaller
attack surface than those with a high fraction of risky services.

\section{Modeling Botnet Infections}
\label{sec:models}

This section examines the effect of the organizational risk vectors on botnet
infections. We start by constructing a model that quantifies how the risk
vectors are correlated with botnet concentration. Next, we consider how these
correlations vary across different industries.

\subsection{Modeling Approach}

We use simple linear regression to model the effect of the various risk vectors
presented in sections~\ref{sec:data} and~\ref{sec:analysis} on botnet
infections. The relationship between peer-to-peer sharing and botnet infections 
appears to be linear in figure~\ref{fig:scatter}.  This suggests
that a linear multiple regression with appropriately transformed variables is a
reasonable method for studying the effect of each risk vector on botnet
infections~\cite{christensen2006log}. 
%We leave the use of more sophisticated models for future work.

\subsubsection{Variable Transformation}

Beginning with the peer-to-peer file sharing data shown in in
section~\ref{sec:analysis}, the apparent linear relationship with botnet
infections a log/log scale suggests a type of power-law relationship 
%between the peer-to-peer file sharing and botnet infections exist. 
To incorporate this variable into our linear regression, we would expect to
log-transform the data first. However, this would entail removing the 13.5\% of
organizations with zero peer-to-peer file sharing and non-zero concentrations
of botnet infections, possibly biasing our results.  To avoid this problem, 
%and to better understand the effect of peer-to-peer file sharing 
we separate the peer-to-peer data into two variables~\cite{hosmer2004applied}:
1) an indicator random variable (binary variable) that is one if the
peer-to-peer file sharing value is zero and 2) a transformed variable
calculated as $log(\text{concentration of peer-to-peer shares})$.
Specifically, if  $t$ is the number of files being shared per employee, we
calculate the indicator random variable, $t_{0}$, as

\begin{equation}
  \label{eq:torrent_zero}
  t_{0} = 
  \begin{cases}
    1,& \text{if } t=0\\
    0,& \text{otherwise}
  \end{cases}
\end{equation}

\noindent and the transformed variable $\hat{t}$ as 

\begin{equation}
  \label{eq:torrent_hat}
  \hat{t} = 
  \begin{cases}
    \log(t),& \text{if } t>0\\
    0,& \text{otherwise}
  \end{cases}
\end{equation}.

This transformation is useful in two ways. First it allows us to incorporate
the observed log/log relationship while including all of data points.  Second,
it allows us to measure both the effect of both presence and prevalence of file
sharing. 

\subsubsection{Regression Model}

Because the rest of the variables presented in section~\ref{sec:analysis} do
not span multiple orders of magnitude and have a roughly linear relationship
with botnet infections, we can include them as linear terms in the regression.
Our final regression model is

\begin{align}
  \label{eq:regression}
  \log (b) =& \beta_1 t_0 + \beta_2 \hat{t} + \nonumber \\
            & \beta_3 T_{CF} + \beta_4 T_{CT} + \nonumber \\ 
            & \beta_5 S_{Ri} + \beta_6 S_{Re} + \nonumber \\
            & \beta_0 + N(0,\sigma)
\end{align}

\noindent where $b$ is botnets per employee, $t_0$ and $\hat{t}$ are the
transformed variables from equations~\ref{eq:torrent_zero}
and~\ref{eq:torrent_hat} respectively; $T_{CF}$ is the fraction of TLS services
with configuration errors; $T_{CT}$ is the fraction of TLS certificates with
certificate errors; $S_{Ri}$ is the fraction of risky services; $S_{Re}$ is the
fraction of reasonable services; $\beta_0$ is the intercept; and $N(0,\sigma)$
is the normal distribution of residuals with mean 0 and standard deviation
$\sigma$.

\subsection{Regression Results}
\label{subsec:regression}

\begin{figure}[t!]
  \includegraphics[width=\columnwidth]{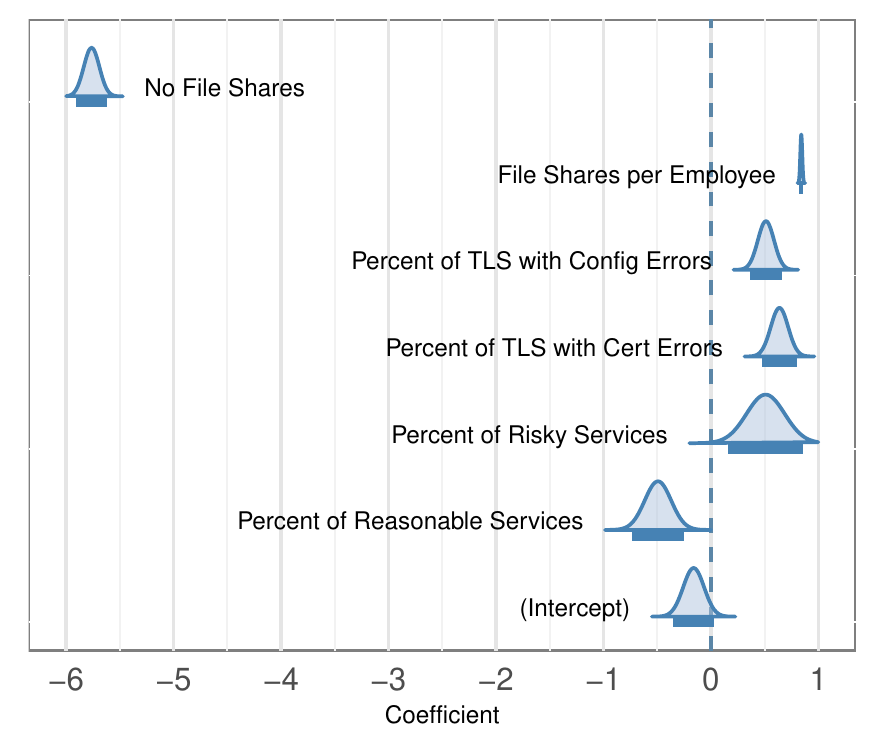}

  \caption{Plot of the regression coefficients from
  equation~\ref{eq:regression}. An estimate of each coefficient's distribution
is given as a Kernel Density estimate on each line. Solid lines below represent
98\% confidence intervals and dark blue lines are significant at $p<0.01$.}
  
\label{fig:pooledBotConcentration}

\end{figure}

We determined the coefficients $\beta_i$ in equation~\ref{eq:regression} using
ordinary least squares. 
%Specifically, we used R's ordinary \texttt{lm} function~\cite{RManual} and
%confirmed the results with Python's \texttt{statsmodels}
%package~\cite{statsmodels}.  We also tested other approaches such as Bayesian
%Generalized Linear models with nearly identical results. 
Post-regression diagnostics indicate that it is unlikely our coefficients
estimates are biased.  The exact coefficients are shown in
table~\ref{tbl:pooledRegression} and depicted visually in
Figure~\ref{fig:pooledBotConcentration}. 

\setlength{\tabcolsep}{2pt}
\begin{table}
  \caption{Estimated Coefficients for the model described
  in~\ref{eq:regression}. All value are statistically significant at the
$p<.005$ level, except for the Intercept which is significant at the $p<0.1$
level.}
  \label{tbl:pooledRegression}
  \centering
  \begin{tabular}{|l|c|c|}
    \hline
    \textbf{Variable} & \textbf{Coefficient} &\textbf{Estimate} \\
    \hline
    Peer-to-Peer Blocked ($t_0$) & $\beta_1$ & -5.763 \\
    \hline
    Peer-to-Peer Concentration ($\hat{t}$) & $\beta_2$ & 0.841 \\
    \hline
    TLS Configuration Errors ($T_{CF}$) & $\beta_3$  & 0.512\\
    \hline
    TLS Certificate Errors ($T_{CT}$) & $\beta_4$ & 0.638 \\
    \hline
    Risky Services ($S_{Ri}$) & $\beta_5$ & 0.509 \\
    \hline
    Reasonable Services ($S_{Re}$) & $\beta_6$ & -0.493 \\
    \hline
    Intercept & $\beta_0$ & -0.167\\
    \hline
    \hline
    \multicolumn{2}{|c|}{$\mathbf{R^2}$} & 0.458\\
    \hline

  \end{tabular}
\end{table}
\setlength{\tabcolsep}{5pt}

The results of the regression indicate that the measured risk vectors affect
the concentration of bots in the expected way.  Organizations with high
concentrations of peer-to-peer sharing, risky services, and configuration and
certificate errors, tend to have higher concentrations of botnet infections. 
%The absence of peer-to-peer sharing and a high percentage of the reasonable
%services in use results in lower botnet concentrations.

Figure~\ref{fig:pooledBotConcentration} illustrates the difference in effect
size between each of the variables. Interestingly, we see that the indicator
variable for peer-to-peer sharing has the largest effect. In particular, we
calculate using the model (equation~\ref{eq:regression}) that organizations
that allow peer-to-peer file sharing have, all other things being equal, more
than 318 times the number of bots per employee\footnote{When the dependent
  variable is log-transformed, a one unit change in an independent variable
results in an $e^{\beta_i}$ change in the dependent variable. Organizations
that block peer-to-peer sharing have $e^{-5.763} = 0.00314$ times lower botnet
infections rate, or inversely $1/0.00314=318$ higher botnet infection rate}. We
explore the reasons for this large effect in the next section and the
section~\ref{sec:discussion}.  We can calculate the effects for other
variables, for example, organizations with 10\% higher peer-to-peer file
sharing per employee have an 8.3\% higher rate of botnet infections.
Organizations with no TLS certificate errors have 14\% lower botnet infections
than those with a 25\% misconfiguration rate.

The variation in the estimates for each variable seen in
figure~\ref{fig:pooledBotConcentration} is also interesting. The narrow
variation in the concentration of peer-to-peer sharing indicates a consistent
effect across different organizations, while the wider variation seen in risky
services indicates an inconsistent effect which perhaps varies significantly
across different organization types.  For example, we hypothesize that the
reason for the weak effect is that in some organizations, risky services might
be necessary and appropriately handled, even if in general this risk vector is
associated with higher levels of botnet infections. We explore how the effect
of different industries influences these relationships in the next section.

%\subsection{Logistic Pooled Regression}}
%
%\begin{figure}[ht!]
%  \includegraphics[width=\columnwidth]{./figs/placeholder.png}
%
%  \caption{Plot of effect size and confidence interval for pooled logistic
%  regression for the presence of bots.}}
%  
%\label{fig:pooledBotPresence}
%
%\end{figure}

\subsection{Industry Effects}
\label{subsec:industry}

\begin{figure}[t!]
  \includegraphics[width=\columnwidth]{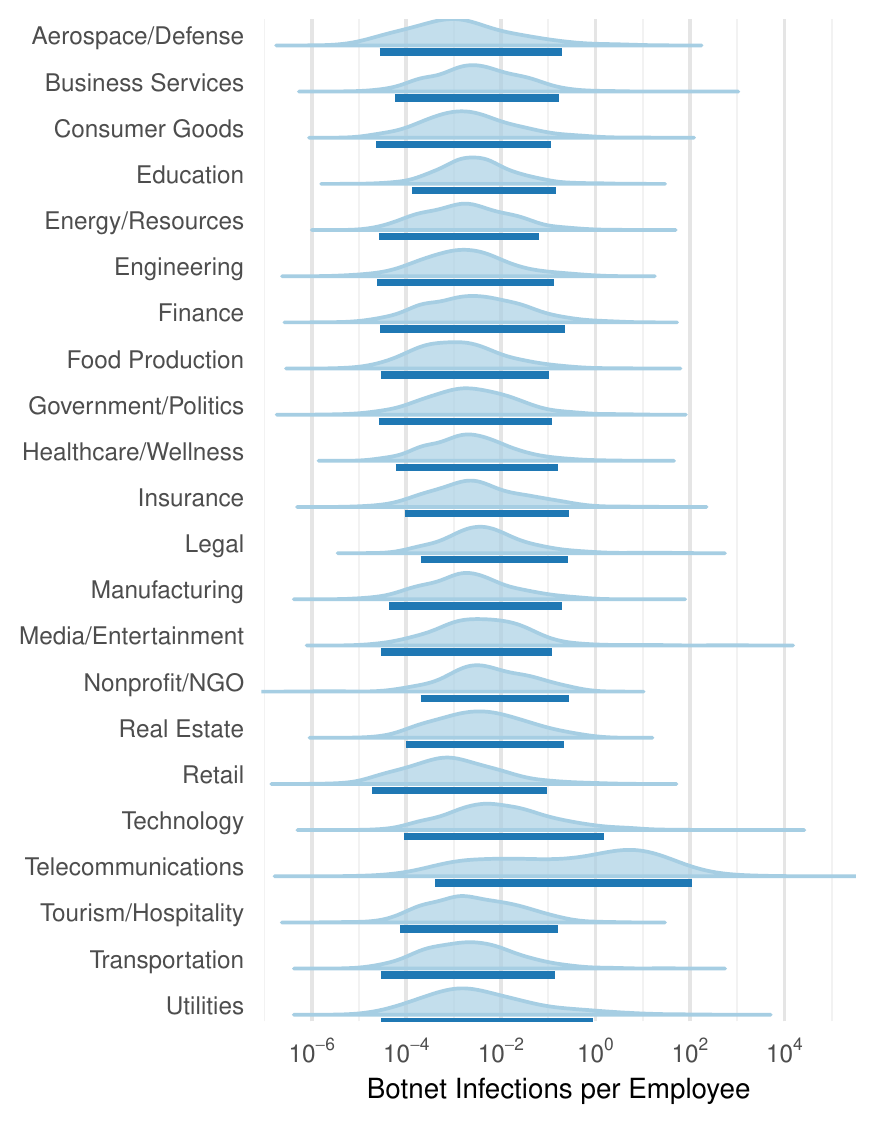}

  \caption{Distribution of botnet concentration by
  industry. Note the log scale, indicating that some industries typically have
many orders of magnitude more botnets per employee than others.}
  
\label{fig:botnet-by-industry}

\end{figure}

Next, we ask whether or not the risk vectors have different effects in
different industry types.  For example, telecommunications companies and
universities typically have less control over the computers connected to their
networks and this may lead to higher botnet infection rates. This effect can be
see in figure~\ref{fig:botnet-by-industry}, where the distribution of botnet
infections for various industries is plotted. As a more rigorous test, we
conducted a pairwise comparison of botnet infection distributions between all
industries using a Kolomogorov-Smirnov test~\cite{smirnov1948table}. Among the
231 possible industry comparisons, 63\% indicated that botnet distributions
between industries differed. We discuss more refined groupings of industries in
section~\ref{sec:discussion}.

\begin{figure*}[t!]
  \includegraphics[width=\textwidth]{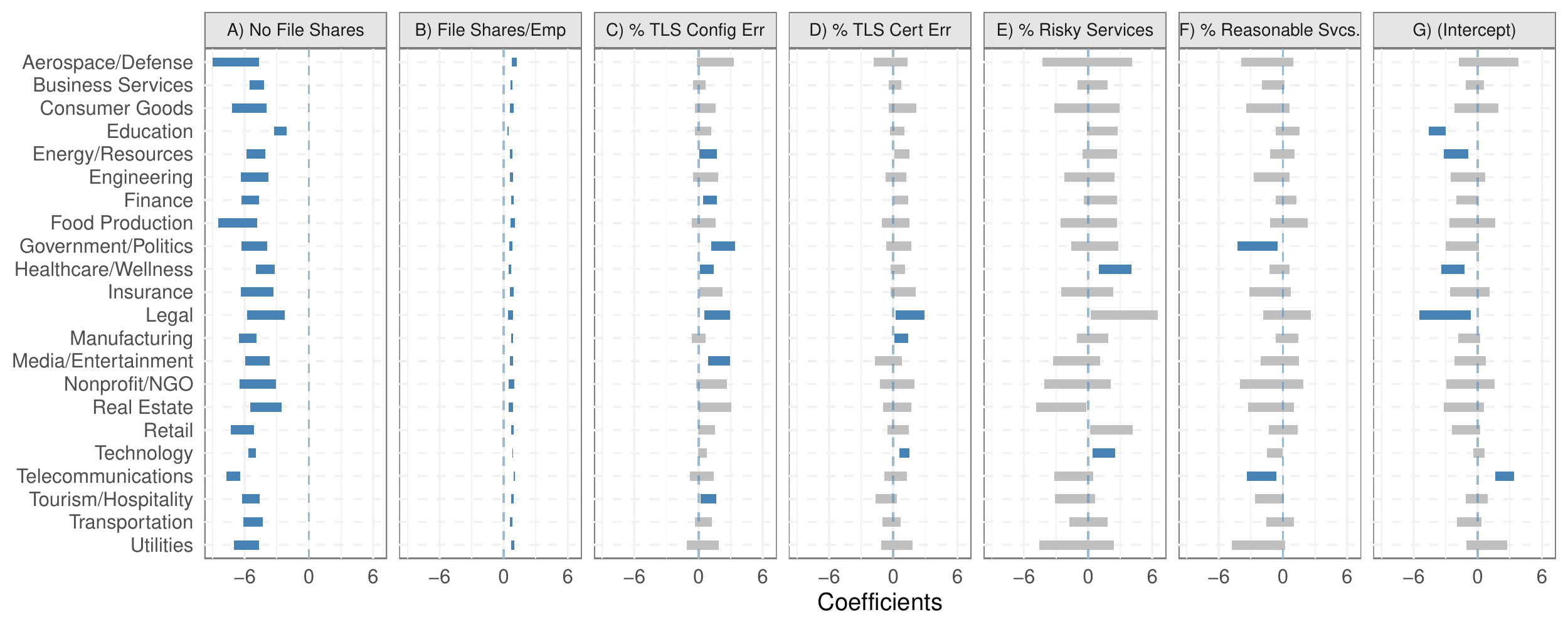}

  \caption{Visual representation of coefficients for the un-pooled
    regression model. Each panel (A-G) shows the coefficients for each
industry in the vertical axis. Each bar represents the 98\% confidence interval
around the coefficient estimate. Estimates which are statistically significant
at the $p<0.01$ level are blue. Non-statistically significant estimates are in
grey.}
  
\label{fig:UnpooledResults}
\end{figure*}

This suggests that the model in equation~\ref{eq:regression} captures only some
of the relevant behavior and that more detailed industry-specific models may be
appropriate.  There are several possible approaches that could be used, e.g.,
we could include an indicator random variable for each industry (fixed effect
model), we could allow a different intercept for each industry (random effect
model), or we could refit the regression model once for each industry (an
un-pooled model). Each of these approaches will result in a more accurate
representation of the data, but at the expense of potentially unnecessary
complexity. We use Akaike Information Criteria (AIC)~\cite{burnham2003model} to
select among different possible models  AIC considers the model's goodness of
fit and penalizes its complexity.  We found that the most complex model, a
separate regression for each industry, gives the minimum (preferred) AIC
despite the large number of variables (154). The results are shown in
figure~\ref{fig:UnpooledResults}\footnote{For brevity, we do not give the exact
values of all 154 coefficients and their standard errors here; however, we will
make them available before publication.}. 

The above plot shows how the effect of risk vectors varies across
industries.  The only consistent effect across all
industries is peer-to-peer file sharing (Columns A and B). 
%and the total
%concentration of peer-to-peer file sharing. 
This indicates that the large
effect observed in section~\ref{subsec:regression} is not confined to
organizations with a large number of hosts per employee, e.g., 
telecommunications and education.  
In particular, Column A has one of the smallest effects 
for educational organizations.  And, Column B
(concentration of peer-to-peer file sharing) is one of
the most consistent across industries, with comparatively low variation
compared to the other coefficients.

Among industries, the effect of the other variables is mixed. For most variables
and most industries, when significant effects are present, they affect botnet
infections in the direction we would expect. TLS configuration, certificate
errors, and risky services all increase botnet
infections, while infection rates decrease as the fraction of safe
services increases.  There is one possible exception to this trend, in
Column E (risky services).
The presence of a large fraction of risky services decreases
botnet infections within the real-estate industry. This effect is not
significant at the $p<0.01$ level but is at the $(p<0.015)$ level. This 
puzzling finding is warrants further investigation.

Finally, for most industries not all risk vectors have a significant effect on
botnet infections. For example, having a high fraction of reasonable
services is significant in only Government/Politics, Business Services, and
Telecommunications industries. Similarly, TLS configuration and certificate
errors only affect a handful of industries. 

\section{Discussion}
\label{sec:discussion}

In this section we highlight some of the more interesting results arising from
our analysis, and we discuss threats to the validity of the results.  We
conclude by exploring a number of possible future research opportunities
suggested by this rich data set and the analysis.

Figure~\ref{fig:violin} showed, counterintuitively, that organizations with
bots tend to have lower concentrations of peer-to-peer file sharing as opposed
to those with no botnet activity.  A second
surprising result is organizations with botnet infections do not have more TLS
configuration errors than those without bots.  Although we do not have a
definitive explanation for these two unexpected results, we speculate that they
could be caused by having set an unrealistic threshold of zero for low botnet
activity.  It is unlikely that over the course of an entire year any
organization would be completely free of botnet infections. Other more
reasonable thresholds for errors may provide a clearer and more intuitive
picture.

%This is a likely
%explanation because when we recompute figure~\ref{fig:violin} using a different
%threshold (figure~\ref{fig:violin2}), the results change in the direction that
%we would have expected.  Specifically, we set the threshold for `low botnet
%activity' to be at or below the 10th percentile.  Selecting the 10th percentile
%as a cutoff is, of course, somewhat arbitrary, and in the future we could take
%a more principled approach to this choice.
%
%\begin{figure}[ht!]
%  \includegraphics[width=\columnwidth]{./figs/bot-nobot-allvars-10pct.pdf}
%
%  \caption{A reproduction of Figure~\ref{fig:violin}, with the threshold for
%  low vs.\ high botnet activity set to the 10th percentile.}
%  
%\label{fig:violin2}
%\end{figure}

Among the risk vectors we studied, the presence and prevalence of peer-to-peer
file sharing through BitTorrent had the largest effect on botnet
concentrations among organizations that showed signs of infection.
Initially we hypothesized that this was an artifact of normalizing
organization size using employee counts rather than number of
computers.  If true, we would expect to see differences in the
un-pooled regression study.  However, peer-to-peer file sharing had a
consistent effect across all industries.
%It is somewhat surprising that this effect has not been
%observed or reported yet, given its magnitude. This may be explained
%by the
Research on file sharing has focused on the performance and
security of the protocol itself, rather than on the security risks it
creates as a side effect (see section~\ref{sec:related}). 

We reiterate that our analysis has identified correlations and that the risk
vectors are not necessarily causal. For example, it is unlikely that an expired
TLS certificate would lead directly to a botnet infection.  However, given the
prevalence of malware present in files shared through
BitTorrent\cite{cuevas2014torrentguard}, it is possible that some of the
infections we measured are the direct result of file sharing.  The most likely
cause for the relationships we see is that both risk vectors and botnet
infections arise from the same common cause: security immaturity. For example,
organizations that don't prevent the use of peer-to-peer file sharing may also
have difficulty identifying and cleaning up botnet infections.

When we created the un-pooled regression model by industry we found only
one surprising result: in the real estate industry, a high fraction of risky
services is associated with lower levels of botnet infections.  This
anomalous result does not have an obvious explanation.
% Case
%studies on the nature of the organizations in this category may reveal a
%clearer picture.

Many of the risk vectors did not have significant statistical effects in the
un-pooled regression.   This could be an artifact of how we defined the risk
vectors originally.  For example, we give equal weight to all TLS errors, but
some errors may be more indicative than others.  A second possibility is that
there are real differences in risks in different industries.  For example,
correct TLS configuration may be less important in the aerospace industry
because its main business does not involve communicating customer data.  In the
future, an organization might use the kind of analysis presented here to
prioritize which security issues to address first.   

\subsection{Caveats and Threats to Validity}

As with any large-scale study, our data might be incomplete or biased.
Variation in monthly scans was small, but even this small amount of variation
may indicate that we are missing some data.
%We detected some variation in the monthly scans, suggesting that we have
%missed some data.
We mitigate this issue by aggregating monthly data into a single yearly measure
for each organization.  In the future, higher resolution data might provide
additional insights. 
%However, rigorous analysis of how is variation over time might affect our
%results will be necessary in the future to fully validate our results.

As discussed earlier, we measure organization size in terms of employee counts,
assuming that this is a good proxy for of the actual number of computers within
the organization.  What we actually care about is the number of computers
connected to each organization's network. However, to our knowledge there is no
straightforward way to obtain such a count.  We know, for example, that
employee count is not particularly accurate for Telecommunications companies
(because employee counts miss all the customers and their computers).  We also
know that in educational institutions, students are not included in employee
counts.
%discussed in section~\ref{sec:models} and can be seen in
%figure~\ref{fig:botnet-by-industry}.
A potential alternative would be to estimate organization by counting the total
number of IP addresses associated with it. We have experimented with this
measure informally with similar results.
%in our mapping and our analysis yielded similar results[WE SHOULD CHECK THIS].

Any measurement that assesses network properties based on IP addresses will be
inexact. The presence of NATs and DHCP and cloud service technology will affect
the accuracy of our mapping.  For example, the extensive use of cloud services
may shrink the perceived network footprint of an organization while inflating
that of the cloud service provider. Network volatility, specifically the
reallocation of IP addresses, also make identifying the exact network
boundaries of an organization difficult.  We did our best to mitigate the
effects caused by these network practices. For example, file sharing was
measured on a per file basis rather than a per IP basis and infections were
measured on an infection per day basis. In both cases our measurements provide
a lower bound on the measurement of interest. In the future, some of these
measurements could be refined. Indirect methods for measuring botnet size have
been proposed~\cite{fabian2007my}, and some botnets utilize unique identifiers
for individual infections which could separate infections on a single IP
address~\cite{stone2009your}. Additionally, data on the internal structure of
organizational networks may provide more accurate measures of risk vector and
infection concentrations.

One of our main results indicates that peer-to-peer communication has a strong
correlation with botnet infections. Previous work has demonstrated that some
botnets have adopted peer-to-peer communication in attempt to grow more
robust~\cite{grizzard2007peer}. It has also been shown that the BitTorrent
protocol itself can be used as a cover channel~\cite{cunche2014asynchronous},
though no study has observed this channel used in a botnet in the wild.
However, our measurement of peer-to-peer file sharing focuses on IP addresses
that are advertising popular files for download, and we believe it is unlikely
that these popular files are being used for covert communication.

We use linear modeling in section~\ref{sec:models}, because our preliminary
analysis (section~\ref{sec:analysis}) indicated that for most risk vectors
linearity (under transformation) was a reasonable assumption.  However, we note
that in figure~\ref{fig:distribution} many of the variables have a multi-modal
distribution, which implies more complexity than we have captured in our linear
regressions. 
%This multi-modal distribution implies a more complex relationship between
%things like TLS configuration errors and bot prevalence. 
However, figure~\ref{fig:scatter} suggests that these extrema are unlikely to
bias our our results. We also explored using multiple variables to represent
these multimodal distributions (similar to the way we handled peer-to-peer
sharing with equation~\ref{eq:torrent_zero}), but models with these transformed
variables did not improve the fit of the model and yielded equivalent results
to those seen in section~\ref{sec:models}.

\subsection{Future Work}
\label{subsec:future}

Future work could address the completeness of the data set by using statistical
techniques to identify the true population size, both for risk vectors and for
outcomes. Mark and recapture methods, originally developed for estimating
species populations in ecology, have already been used to estimate the true
size of botnets~\cite{weaver2010probabilistic}. These methods could be used to
improve our estimates of botnets and risk vectors.

We have simplified some risk vectors, for convenience and ease of
interpretation.  For example, we weight all TLS software errors equally,
although it is unlikely that they are all equally risky.  For example, an
organization that uses the theoretically weak 1024-bit Diffie-Helman key
exchange may not actually experience more attacks.  Similarly, we made a rough
classification of services into risky and reasonable. Services we did not
investigate may be strong indicators of negative organizational outcomes.
These nuances may explain the multi-modal distributions observed in
figure~\ref{fig:distribution}. However, separating out these effects is
challenging because of concerns about statistical independence.  
% as many of the errors
%described to measure mis-configured TLS are highly correlated. 
More complex modeling approaches, such as hierarchical models, could help
identify more precise relationships.

While the current set of risk vectors can explain nearly half the variation we
seen in botnet infection rates ($R^2=0.49$), there are certainly other measures
of security maturity that could be include to give a more complete picture of
security maturity. We analyzed 21 different services and roughly categorized
them into risky, neutral and reasonable. Other services and a finer
categorization may provide more detailed results. Additionally, internal
organizational practices are likely a good measure of security maturity. For
example, password policies, security training, and internal access control
policies are likely to reflect the security maturity of an organization and
correlate with outcomes like botnet infections. Quantifying these properties
and identifying their effect is rich ground for future work.

Although we examined several different methods to account for difference across
industries, there are many more possibilities.
%industries on botnet infections. However, this is only a small percentage of
%the possible models we could investigate. 
Figures~\ref{fig:botnet-by-industry} and~\ref{fig:UnpooledResults} show botnet
infections and the effect of risk vectors vary across industries.
%however, many of those distributions appear similar. 
However, it is possible that better groupings of industries could provide more
insight in how risk vectors affect security outcomes.

In this paper, we focused on botnets as a measure of negative outcomes for
organizations. However, the methodology can easily be applied to other security
problems, such as data breaches or
%
%only measures of botnet activity. Using data that described IPs which had been
blacklists (for sending spam, port scanning, or hosting malware).  Our
preliminary results on these other outcomes are consistent with those presented
for botnets.  %Moreover, our approach could be assess
%how risk vectors affect the likelihood of a data breach.  Liu et al.\ took a
%predictive approach and identified some features which when combined in a
%random forest were able to accurately predict breaches~\cite{liu2015cloudy}.
%However this work did little to identify the exact effect of external risk
%vectors.

\begin{figure}[t!]
  \includegraphics[width=\columnwidth]{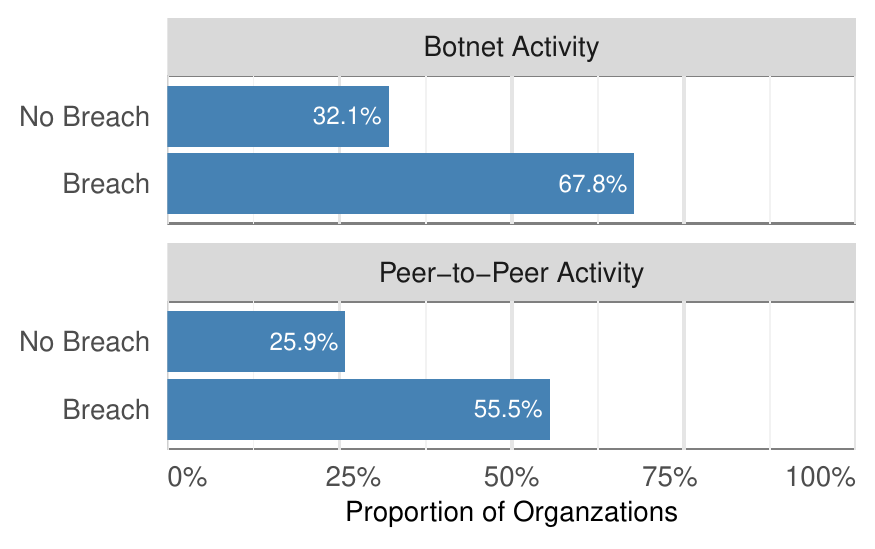}

  \caption{Percentages of organizations which experienced breaches with and
  without botnet activity and peer-to-peer activity.}
  
\label{fig:breaches}

\end{figure}

As one example of generalizing to other problems, we examined a small set of
data breaches experienced by the organizations in our study.  The data were
collected in a variety of ways, such as scraping news articles, Freedom of
Information Act requests, and some private data streams.
%\footnote{We leave full analysis of breach data for future work}. 
We considered two possible risk factors (peer-to-peer or botnet
activity\footnote{Although botnet activity is an outcome in the earlier
sections, here we treat it as a risk factor itself.}), and found that
organizations with data breaches are likely to also have these risk factors
(Figure~\ref{fig:breaches}). This effect is statistically significant according
to the G-test ($p<10^{-12}$).   This preliminary work is promising but requires
additional study and validation before we can draw firm conclusions.

\section{Related Work}
\label{sec:related}

In this section we review work related to that presented in this paper. We find
that research as focused on identifying vulnerabilities and elaborating on why
they might put organizations at risk, there has been surprisingly little work
linking risk vectors to actual outcomes such as botnet infections. 

Work on BitTorrent has primarily focused on measuring the use of the protocol
as a method for peer-to-peer file distribution~\cite{pouwelse2005bittorrent},
it's performance~\cite{bharambe2005analyzing}, the security of its distributed
hash table algorithm~\cite{timpanaro2011bittorrent}, and studies of attacks
against the protocol~\cite{konrath2007attacking,hatahet2010new}. However, there
has been little work studying the potential danger of the use of the protocol
in the wild. To our knowledge, only one paper by Cuevas examined the dangers of
BitTorrent identifying that 35\% of files shared on through BitTorrent are
fake, and of those more than 99\% contain malware or phishing
attempts~\cite{cuevas2014torrentguard}. Our work is the first which actually
examine this correlation between the use of peer-to-peer file sharing in
organizations and botnet infections. 

In contrast, the Transport Layer Security protocol, and its predecessor the
Secure Socket Layer protocol, have been widely known to be a source of
vulnerabilities. One recent and widely publicized was the ``Heartbleed''
vulnerability which allowed attackers to remotely read protected memory on
servers hosting vulnerable websites~\cite{durumeric2014matter}. Other research
has shown that many of the available cryptographic protocols used to
communicate, and hash functions used to sign certificates are vulnerable to a
variety of attacks~\cite{beurdouche2015messy,
adrian2015imperfect,FoxIT2011,wang2005finding,dobbertin1996status,rogier1997md2}.
There have also been work suggesting that the certificate infrastructure itself
is a target for hackers, in particular one of the major certificate
authorities, Verisgn, was the subject of numerous successful attacks in
2010~\cite{menn2012key}. While this work speaks to the insecurity of various
implementations and configurations of TLS, and the certificate infrastructure
in general, it does provide a direct link between the problems of TLS and
outcomes for organizations. Reports have linked exploitation of the Heartbleed
bug to the compromise of 4.5 million patient records in an attack against
Community Health Systems~\cite{leyden2014heartbleed}, an unknown number of
records against the website ``Mumsnet''~\cite{quinn2014parenting}, and a
compromise of an unknown number of records from Canadian Revenue
Agency~\cite{seglins2014CRA}. However, no work to our knowledge has
investigated the relationship between vulnerable TLS services and botnet
infections.

The case is similar for various types of services. As we outline in
section~\ref{sec:data}, there has been extensive research on developing attacks
for a number of the protocols we collect data on including
FTP~\cite{manadhata2006measuring}, TELNET~\cite{joncheray1995simple}, and
Microsoft SQL~\cite{cerrudo2002manipulating}. Moreover, techniques have been
developed for automatically generating attacks against a variety of
protocols~\cite{guruswamy2011portable}. However, as of now we find no work
linking the presence or use of these services to negative outcomes, as we do in
this paper.

Botnets are recognized as a costly part of the cybercriminal
infrastructure~\cite{anderson2013measuring}, and a good deal of research has
been devoted to detecting~\cite{feily2009survey},
measuring~\cite{abu2006multifaceted,dagon2006modeling},
classifying~\cite{dagon2007taxonomy} and fighting
them~\cite{van2010role,bailey2009survey}. Similar to the research presented in
this paper, some work focuses on measuring relative botnet infections within
various types of organizations. For example Stone-Gross et al.\ focused on ISPs
with persistent malicious behavior~\cite{stone2009fire}.  Edwards et al.\
studied the concentration of spam sending IP addresses within Internet Service
providers, and examined some risk vectors(including economic, geographic, and
connectivity) for high levels of spam
concentrations~\cite{edwards2015analyzing}. Other work has focused identifying
high concentrations of infected IP addresses in certain parts of the
Internet~\cite{moura2013internet,ramachandran2006understanding,collins2007using}.
Yen et al.\ explored malware infections in a single large enterprise, and
identify possible entry points for infection on individual hosts, but do not
provide a comparison across a large number of organizations as we do
here~\cite{yen2014epidemiological}.

Little work linking the network characteristics of organizations with security
incidents across a broad number of organizations. Zhang et al.\ found that
higher concentrations of different types of network mismanagement such as open
DNS resolvers and SMTP relays lead to high concentrations of infected PCs
within different Autonomous Systems~\cite{zhang2014mismanagement}. Our work
focuses on an organizational level, and considers a broader scope of risk
vectors which are correlated with network infections. Liu et al.\ use a similar
but much larger set of network features (258) of organizations and machine
learning to predict data breaches~\cite{liu2015cloudy}. While they make some
effort to rank features on their relative importance, using a random forest
limits their ability to directly quantify the impact of risk vectors as we do
here.

In the past there have been calls to better understand security at the
organizational level~\cite{whitman2003enemy}. However, research developing
qualitative models of organizational security have been difficult to
validate~\cite{kotulic2004there,siponen2000conceptual}. Others have tried to
investigate organizations security culture. Merete et al.\ study common
information security practices across organizations in
Norway~\cite{merete2008implementation}, and Furnell et al.\ investigate how
organizational culture can be utilized to increase
security~\cite{furnell2005organizational}. Recent work has tested organizations
response to being told informed about spam sending infections, both publicly
and privately~\cite{he2015designing}. However, none of this work makes a direct
connection to organizational maturity and the likelihood of incident such as
botnet infections.

\section{Conclusions}
\label{sec:conclusion}

The ability to assess security risks in organizations is critical,
both internally and externally.  Internally, those who are most
responsible for a company's performance, e.g., the CEO or Board of
Directors, often lack the technical skills to assess their own IT
systems, especially for security risks.  The evidence-based approach
described in this paper provides an objective and quantitative method
that goes beyond self-reporting or qualitative audits.  From an
external perspective, these methods could be used by one organization
to guide decisions about outsourcing or partnerships, and they could
potentially be used in settings like insurance where quantifying risk
is essential.  Current approaches provide only the most general idea
of exactly what practices and vectors expose organizations to minor
incidents such as malware infection or major incidents such as data
breaches, business interruptions, and financial and intellectual
theft.  Although the results in this paper represent a first cut at
developing a robust quantitative approach to assessing security risks,
the methods could readily be applied to other data sources, other risk
vectors, and other security problems.

Even as a first cut, our results are promising. We found that 90\% of
organizations have low levels of botnet infections (fewer than 1 botnet
infection per 12 employees) however, infection rates can span many orders of
magnitude.  Using a simple, linear regression approach, we found that the
presence and prevalence of peer-to-peer file sharing through the BitTorrent
protocol, TLS errors, and publicly available services are all correlated with
concentrations of botnet infections. In particular we find that organizations
which have peer-to-peer activity have, on average, 318 times higher
concentrations of bot. When we evaluate different industries separately, a
similar large effect was consistently observed all different industries;
however, other risk vectors are only significantly related to botnet infections
in a handful industries.

%As our reliance on technology and the
%scope of cyber-insecurity increases, organizations will need more than the
%rough estimates provided by today's risk assessment questionnaires and audits.
%A more sophisticated approach is needed. 

In conclusion, we argue that a data-driven, statistically principled approach
is the best way to identify objective risks within an organization. Identifying
what risk vectors and minor incidents are correlated with major incidents such
as data breaches is an important next step in understanding security maturity. 
%This data-driven approach to security will help guide organizations and policy
%makers towards evidence-based decision making. 

\section{Acknowledgements}

The authors would like to thank BitSight for providing access to their data to
make this research possible.

\bibliographystyle{plain}
\bibliography{paper}

\end{document}